\def \ber{\begin{eqnarray}}
\def\eer{\end{eqnarray}}

\documentclass[aps, jcp, prb,  two column,showpacs,superscriptaddress]{revtex4-1}
\usepackage{float}
\usepackage{graphicx} 
\usepackage{subfigure}
\usepackage{dcolumn}
\usepackage{multirow}
 \usepackage{bm}      
\usepackage{amssymb,amssymb}   
\usepackage[left= 1 in, right= 1 in, top= 1 in , bottom= 1 in]{geometry}
\usepackage{hyperref}
\usepackage{xcolor} 
\hypersetup{ colorlinks, citecolor=blue ,linkcolor=black }
\begin{document}
\title{ Transport properties of Methane, Ethane, Propane, and n-Butane in Water}
\author{Sunil Pokharel}
\email{physicistsupo@gmail.com}
\affiliation{Central Department of Physics, Tribhuvan University, Kirtipur, Kathmandu, Nepal}
\author{ Narayan Aryal}
\email{aryaln65@gmail.com}
\affiliation{Central Department of Physics, Tribhuvan University, Kirtipur, Kathmandu, Nepal}
\author{ Bhakta Raj Niraula}
\email{brajnc246@gmail.com}
\affiliation{Central Department of Physics, Tribhuvan University, Kirtipur, Kathmandu, Nepal}
\author{Arjun Subedi}
\email{arjubedi@gmail.com}
\affiliation{Central Department of Physics, Tribhuvan University, Kirtipur, Kathmandu, Nepal}
\author{ Narayan Prasad Adhikari}
\email{npadhikari@gmail.com, npadhikari@tucdp.edu.np}
\affiliation{Central Department of Physics, Tribhuvan University, Kirtipur, Kathmandu, Nepal}
\date{\today}

\begin{abstract}
\noindent  In this work, we have estimated self diffusion coefficients along with the binary diffusion 
coefficients of mixtures of alkane  (methane, ethane, propane and n-butane) in SPC/E water(H$_2$O). 
Molecular dynamics study of a binary mixture of alkane gas and SPC/E water, with alkane as solute 
and water as solvent, have
been carried out at different temperatures ranging from 283.15 K to 333.15 K. We have taken a dilute  
solution of 3 alkane (methane, ethane, propane and n-butane) molecules and 971 water molecules in a system. 
The role of interaction in the structure of the constituents of the system  as a function of temperature 
is studied with the help of the radial distribution function (RDF) and the coordination numbers. 
The self-diffusion coefficient of the constituents of the mixture  was calculated by using mean 
square displacement (MSD) and the binary diffusion coefficients of alkane in water have been calculated by
using Darken's relation. The results are then compared with the available experimental values. 
The  values of self-diffusion coefficients of water from the present work  come in good agreement 
with the experimental values  within 9$\%$ error.  The binary diffusion coefficients of ethane, methane, propane 
and n-butane agree with the previously reported experimental values. 
The  dependence of the diffusion coefficients on  temperature is  approximated by Arrhenius-type exponential relationship.\\\\
   
\end{abstract}
\date{\today}
\maketitle
\section{Introduction}
\noindent
Alkanes are saturated hydrocarbons that consist only of the elements carbon (C) and hydrogen (H), where each of these atoms are linked together exclusively by single bonds. Alkanes belong to a homologus series of organic compounds in which the members differ by a constant molecular mass of 14 that is $\mathrm{CH_2}$ \cite{Morrison}. The smaller members of the alkane family are gases, while the larger are liquid and solid compounds. The alkanes are soluble in non-polar solvents such as benzene, ether and chloroform, and are insoluble in water and other highly polar solvents. The most important sources of alkanes are natural gas and crude oil. Petroleum and natural gas are largely mixtures of different alkanes. On refining, they give liquefied petroleum gas (LPG),
gasoline, kerosene, diesel, furnace oil and wax which are used as fuels. The solid alkanes (compounds) are typically waxy in texture. The uses of alkanes can be determined according to the number of carbon atoms present in it. Some of the common uses of alkanes are heating, electricity generation, cooking, production of polymers, serves as intermediate in the synthesis of drugs, pesticides and other chemicals,  components of gasoline (pentane and octane) and paraffin wax. First four members of alkane series are methane, ethane, propane, and butane with molecular formula $\mathrm{CH_4}$, $\mathrm{C_2H_6}$, $\mathrm{C_3H_8}$, and $\mathrm{C_4H_{10}}$ respectively. The first four alkanes are used for heating, cooking and electricity generation. The main components of natural gas are methane and ethane. Propane and Butane are used as LPG (liquified petroleum gas). Propane is also used in the propane gas burner, butane in disposable cigarette lighters. They are also used as propellants in aerosol sprays. The processing of natural gas involves removal of propane, butane, and large amount of ethane from the raw gas in order to prevent condensation of these volatiles in natural gas pipelines Alkanes have a number of industrial applications beyond fuels, including uses in cosmetics and plastics. Alkanes are generally less reactive than alkenes and alkynes because they lack the more reactive double and triple bonds. However, they do participate in reactions with oxygen (combustion) and halogens\cite{Morrison, Arora, Roberts, Weber}.\\
\noindent Modeling and Computer simulation are carried out in the hope of understanding the properties of assemblies of molecules in terms of their structure and the microscopic interactions between them. This serves as a complement to conventional experiments, enabling us to learn something new, something that cannot be found out in other ways \cite{frenkel, allen, berendsen, akira, rapport, becker}. The two main families of computer simulation  technique are molecular dynamics (MD) and Monte Carlo (MC); additionally, there is a whole range of hybrid techniques which combine features from both. The obvious advantage of MD over MC is that it  gives a route to dynamical properties of the system: transport coefficients, time-dependent
responses to perturbations, rheological properties and spectra \cite{frenkel, allen}\\
\noindent Molecular Dynamics simulation is a technique for computing the equilibrium and transport properties of a classical many-body system. In this context, the word classical means that the nuclear motion of the constituent particles obeys the laws of classical mechanics. This is an excellent approximation for a wide range of materials. Molecular dynamics simulation consists of the numerical, step-by-step, solution of the classical equations of motion \cite{frenkel, allen}
\begin{equation}
m_i\frac{d^2\mathbf{r}_i}{dt^2} = \mathbf{F}_i = -\nabla_{\mathbf{r_i}}\mathrm{ U }
\end{equation}
For this purpose we need to be able to calculate the forces $\mathbf{F}_i$ acting on the atoms, and these
are usually derived from a potential energy $U(\mathbf{r}^N)$, where $\mathbf{r}^N = (\mathbf{r}_1, \mathbf{r}_2, . . .,\mathbf{r}_N)$ represents the complete set of 3N atomic coordinates.\\

\noindent  As different species of a mixture move under the influence of concentration inhomogeneity, molecular diffusion occurs \cite{diffusion in solids}. It plays a vital role in variety of biospheric and atmospheric sciences. Diffusion is basic for transport of matter and for ionic conduction in disordered materials\cite{na-sir1, na-sir2}. The kinetics of many micro structural changes that occur during preparation, processing and heat treatment of materials include diffusion. The typical examples are nucleation of new phases, diffusive phase transformation, precipitation and dissolution of a second phase, homogenization of alloys, recrystallization and thermal oxidation
\cite{diffusion in solids}. Alkanes are less reactive in comparison to other chemical species because carbon
atoms in alkanes have attained their octet of electrons through forming four covalent bonds. These four bonds formed by carbon in alkanes are called sigma bonds, which are more stable than other types of bond because of the greater overlaps of carbon’s atomic orbitals with the neighbouring atom’s atomic orbitals. Alkanes are non-polar solvents as only   $\mathrm{C}$ and $\mathrm{H}$ atoms are present. Alkanes are insoluble in polar solvent like water but freely soluble in non-polar solvent like ether and benzene. Since alkanes are rarely soluble in water, it is very appropriate and interesting to study the diffusion of alkane in water. The first four members of alkane series exist in the vapour state under the normal atmospheric condition, which are principle ingredients in natural gas. Melting point and boiling point of alkanes increase with the increase in number of carbon atoms. The principle source of alkanes is petroleum, which consists together with the accompaining natural gas. The complicated organic compounds that once made up living plants or animals have transformed into a mixture of alkanes ranging in size from one
carbon to $30$ or $40$ carbons by decay and millions of years of geological stresses \cite{Morrison}.\\

\begin{figure}[H]
\includegraphics[scale=0.48]{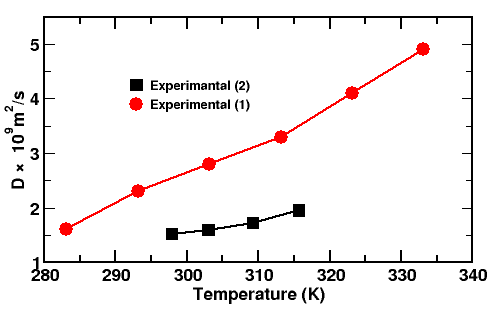}
\caption{Plot of binary diffusion coefficient of ethane-water system  versus temperature of two experimental works \cite{Houghton, Witherspoon}}.
\label{expdiffusion}
\end{figure}

It is found that the diffusion of hydrocarbons (alkanes) in water is a basic consideration in many processes like processing of natural gases and petroleum,  understanding the tertiary structure of proteins, as well as the important role it plays as a driving force in a number of processes occurring within living cells. The experimental values of binary diffusion coefficients of alkane-water mixture have been obtained by D.~L.~ Wise and G.~ Houghton by  using the rate of collapse of small bubbles in gas free water\cite{Houghton}. Also, P. A. Witherspoon and D. N. Saraf have been obtained by using the capillary cell method \cite{Witherspoon}. The result of both techniques are different. Figure (\ref{expdiffusion}) shows the plot of the binary diffusion coefficients of the two experimental works. The  value of binary diffusion coefficients from \cite{Houghton} are deviated at most  76 \% than that of \cite{Witherspoon}.  Such discrepancies in these works motivated us to carry out a computational work to study the diffusion phenomena of alkane in water. Our results obtained from simulation also can be used as a crude reference for any further studies of diffusion in complex  fluid mixtures  and  improve our understanding of hydrocarbons  and other more complex biological macromolecules like protein  in water\cite{protein}.

\noindent The outline of the paper is as follows: In Sec. II, we discuss the theory of  diffusion  and method of calculation of diffusion coefficient. Computational details of the our work are stated in  Sec. III. Results of the our  work are  presented in Secs. IV. Our conclusions are collected in Sec. V.

\section{ Diffusion Coefficient} \label{diffusion_coefficient}
\noindent Diffusion is the process by which matter is transported from one part of the system with higher concentration to the another part of the system with the lower concentration as a result of random molecular motion. The driving force of diffusion is thermal motion of the molecules. The higher concentration of a species in a
system at a particular site corresponds to its higher value of chemical potential. The net transport of the mass takes place from the region of higher chemical potential
to the region of lower chemical potential. At the end of such net transport, the system attains a situation where there exists same value of chemical potential in the
system. In this situation, free energy of the system is minimum and hence its entropy becomes maximum; system is at dynamic equilibrium \cite{Kittel}. The response property of a system to a concetration gradient  is measured by diffusion coefficient \cite{frenkel}. The diffusion in a homogeneous system where no chemical concentration gradient exists is known as self-diffusion  and the corresponding diffusion coefficient is called self-diffusion coefficient \cite{SP}. There are two common ways to obtain a self-diffusion coefficient. The mathematical expression to calculate self-diffusion coefficient from molecular positions is famously known as Einstein relation \cite{frenkel, allen}.
For 3-D system,
\begin{equation}\label{MSD_equation}
D= \lim_{t \rightarrow \infty}\frac{\left\langle [\textbf{r}_{\alpha}(t+t_0) - \textbf{r}_{\alpha}(t_0)]^2 \right\rangle}{6\; t}
\end{equation}
\noindent where $\alpha$ denotes the type of component (solute or solvent) and $t_0$ is any time origin. The angled brackets
$\left\langle \ldots \right\rangle$ indicate the ensemble average. The ensemble average is taken over all atoms of the component $\alpha$ in
the simulation and all time origins \cite{formaldehyde}. The method using Einstein relation for calculating diffusion coefficients is known as MSD  method.\\
\noindent In this work, we calculate the self diffusion coefficients of both the components i.e. alkane (methane, ethane, propane, n-butane) and  water (H$_2$O) which can be used to estimate the binary diffusion coefficient using Darken's relation \cite{darken}
\begin{equation}\label{darken_equation}
\mathrm{D}_{\mathrm{AB}} = \mathrm{N}_\mathrm{B}~\mathrm{D}_\mathrm{A} + \mathrm{N}_\mathrm{A}\;\mathrm{D}_\mathrm{B}
\end{equation}
\noindent where D$_\mathrm{A}$, D$_\mathrm{B}$ are the self diffusion coefficients of species A and B respectively and
N$_\mathrm{A}$, N$_\mathrm{B}$ are the corresponding mole fractions.
\section{Computational Details}
\subsection{Molecular Models}
\noindent The SPC/E ( simple point charge/extended) potential  model \cite{SPCE} is used in all the simulation for water as a solvent. The OPLSS-AA (Optimized Potentials for Liquid Simulations-All Atom) potential model \cite{OPLS}  is used for alkanes (methane, ethane, propane, n-butane) as  solute.  The system under study consists of $3$ alkane (methane, ethane, propane, n-butane) molecules and 971 water molecules separately. In classical force fields like OPLS-AA, the potential functions are derived empirically to describe the atomic interactions. The atoms are treated as spherically symmetric particles and are considered to be connected through covalent bonds to form molecules. Each and every atom experiences a force resulting from its pairwise additive interactions with the rest of the system. The total potential energy $\mathrm{U_{tot}}$  includes
contributions from both bonded and  non-bonded interactions  \cite{gromacs user manual}. The bonded interactions are bond stretching (2-body), bond angle (3-body) and dihedral angle (4-body) interactions. A special type of dihedral interaction (called improper dihedrals) is used to force atoms to remain in a plane or to prevent transition to a configuration of opposite chirality (a mirror image). The non-bonded   interactions are
represented by the van der Waals potential  and Coulomb potential. Therefore, the  total potential energy
function of a system can be written as  as \cite{gromacs user manual}:
\begin{eqnarray}
\mathrm {U_ {tot}} =\mathrm {U_ {b}} +\mathrm {U_ {nb}} \hspace{3.8cm} \\ 
\mathrm {U_ {tot}} = \mathrm {U_ {b}} + \mathrm {U_ {a}} + \mathrm {U_ {d}} + \mathrm {U_ {id}} + \mathrm {U_ {vdw}}  + \mathrm {U_ {c}}
\end{eqnarray}
The bond stretching between two covalently bonded atoms $i$ and $j$ is represented by harmonic
potential \cite{gromacs user manual}
\begin{equation}
U_b(r_{ij}) = \frac{1}{2} k_{ij}^b (r_{ij} - b_{ij})^2
\end{equation}
\noindent where $k_{ij}^b$ is the force constant and b$_{ij}$ is the equilibrium bond length between two atoms $i$ and $j$. The bond angle vibration
between a triplet of atoms $i-j-k$ is also represented by a harmonic potential on the angle $\Theta_{ijk}$ \cite{gromacs user manual}
\begin{equation}
U_a(\Theta_{ijk}) = \frac{1}{2} k_{ijk}^\Theta (\Theta_{ijk} - \Theta^0_{ijk})^2
\end{equation}
\noindent where $k_{ijk}^\Theta$ is the force constant and $\Theta^0_{ijk}$ is the equilibrium bond angle.\\
The proper dihedral angle is defined by the angle between the $ijk$ and $jkl$. The periodic dihedral potential is defined by \cite{gromacs user manual} :
\begin{equation}
\mathrm {U_ {d}} =\mathrm{k^{\phi}_{ijkl}} (1 + Cos (n_{ijkl} ~\phi_{ijkl}-\phi_{0} ))
\end{equation}
Here $\phi_{ijkl}$ is dihedral angle, $\mathrm{k^{\phi}_{ijkl}}$ is the the force constant, $\phi_{0}$ is the reference angle where the potential passes through its minimum value, and $n_{ijkl}$ is the multiplicity, which indicates the number of minima as the bond is rotated through $2\pi$. The multiplicity is a nonzero, positive integer number.\\
For alkanes, the following proper dihedral potential (Ryckaert-Bellmans function) is often used\cite{gromacs user manual}: 
\begin{equation}‎
\mathrm{U_{rb}} =\sum \limits_{n=0}^5    C_n ~(Cos(\psi))^n
\end{equation}
where $\psi = \phi -180^{\circ}$.\\
 The improper dihedral potential is harmonic potential whic is given by \cite{gromacs user manual}: 
 \begin{equation}
 \mathrm{U}_{id} = \frac{1}{2} k_{\xi}~ (\xi_{ijkl}- \xi_{0})^2
 \end{equation}
 where the parameters $\xi_{0}$ and $k_{\xi}$ mark the equilibrium improper dihedral angle and force constant respectively.\\
\noindent The non-bonded interatomic interaction is the sum of Lennard- Jones interaction and Coulomb interaction. The interatomic interaction thus can be written as
\begin{equation}
U_{\alpha\beta}(r_{ij}) = 4\epsilon_{\alpha \beta}\left[\left(\frac{\sigma_{\alpha\beta}}{r_{ij}}\right)^{12} - \left(\frac{\sigma_{\alpha\beta}}{r_{ij}}\right)^6\right] +  \frac{q_{i\alpha}\;q_{j\beta}}{4\pi\epsilon_0\;r_{ij}}
\end{equation}
\noindent where $r_{ij}$ is the Cartesian distance between the two
atoms $i$ and $j$; $\alpha$ and $\beta$  indicate the type of the
atoms. The parameters for the non-bonded Lennard Jones interaction between two different atoms for OPLS-AA force field  are written as  \cite {gromacs user manual}: 
\begin{eqnarray} 
\mathrm{\sigma}_{\alpha\beta} =  \left(\mathrm{\sigma}_{\alpha\alpha} \times \mathrm{\sigma}_{\beta\beta} \right)^{\frac{1}{2}}\\
\mathrm{\epsilon}_{\alpha\beta} = \left(\mathrm{\epsilon}_{\alpha\alpha} \times \mathrm{\epsilon}_{\beta\beta} \right)^{\frac{1}{2}} 
\end{eqnarray}
\subsection{Simulation Set up}
\begin{figure}[H]
\includegraphics[scale=0.45]{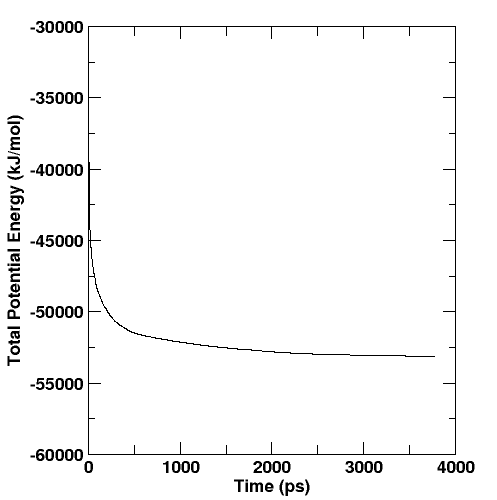}
\caption{Plot of potential energy as a function of time after energy minimization for methane-water system.}\label{enm}
\end{figure}

\begin{figure}[H]
\includegraphics[scale=0.39]
{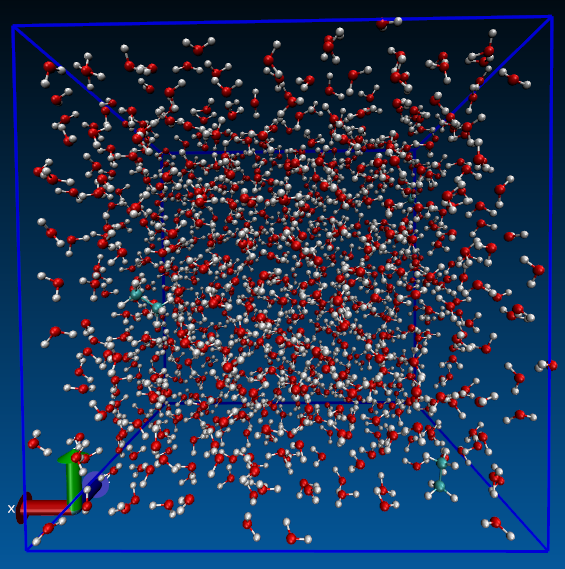}
\caption{Structure of ethane-water system afterequilibration} \label{structure}
\end{figure}

 \noindent MD simulation was carried out in a cubic box  with periodic boundary conditions \cite{allen} using \textbf{GROMACS 4.6.5}. The distance to the edge of the box from the solute (alkane) is an important parameter for defining the size of the box. Since we are  using periodic boundary conditions, we must satisfy the minimum image convention. That is alknane (solute) should never see its periodic image, otherwise the forces calculated will be spurious. The size of the box defined here  is sufficient for just about any cutoff scheme commonly used in simulations. After solvation, addition of 971 water molecules and 3 alkanes molecules in simulation box,
energy minimization is carried out with a cut off restriction of 1.0 nm to avoid unphysical van der Waals contact caused by the atoms that are too close \cite{gromacs user manual}. Energy minimization brings the system to equilibrium configuration, removes all
the kinetic energy from the system, reduces thermal noise in structure and brings the system to one of the local minimum. Steepest descent 
algorithm has been used for energy minimization and the algorithm stops when the maximum of absolute value of force components is smaller
than the specified value \cite{gromacs user manual}. The energy (potential) of the system after energy minimization is shown in figure (\ref{enm}).
\noindent After energy minimization, equilibrium was carried out at different temperatures, 283.15 K, 293.15 K, 303.15 K, 313.15 K, 323.15 K, 333.15 K and a pressure of $10^5$ N$m^{-2}$ (i.e.~NPT Ensemble) by using \emph{velocity-rescaling} thermostat and Berendsen barostat \cite{gromacs user manual} at a coupling time $\tau_t$ = 0.01 ps and $\tau_p$ = 0.8 ps respectively. Here, the system is subjected to NPT ensemble to bring the parameters like temperature, pressure, density, etc. to thermodynamic equilibrium because dynamic property like diffusion coefficient varies with
such parameters. We used MD integrator with time step size 0.002 ps for  $10^9$ steps, which makes equilibration run of 200 ns. The velocity is generated initially according to a Maxwell distribution function at a specified temperature \cite{gromacs user manual}. All the bonds are converted to constraints using SHAKE algorithm \cite{gromacs user manual}. During equilibration short range coulomb interaction and Lennard Jones interaction each with a cut off parameter of 1.0 nm were considered  with periodic boundary conditions\cite{allen}. The long range Coulomb interaction is handled via the PME (Particle Mesh Ewald) algorithm with fourier spacing 0.12. The input parameters (force field parameters, and coupling constants for barostat and thermostat) were taken so as to be consistent with the experimental values as much as possible. The structure of the system after equilibration is shown in figure(\ref{structure}). The density and simulated temperatures at different coupling temperatures for propane in water are shown in table (\ref{density_box}).
 
\begin{table}[H]
\caption{Values of simulated temperature ($T_{sim}$) and density at various coupling temperatures ($T_{co}$) of propane-water system.}
\label{density_box}
\resizebox{0.50\textwidth}{!}{%
\begin{tabular}{c c c c c}\hline
S.N &($T_{co}$) K & ($T_{sim}$) K & $\rho_{sys}$($kg/m^3$) & $\rho_{w}$($kg/m^3$)
\cite{SP, Poudyal2014}\\  \hline
1 & 283.15 & 283.144$\pm$0.005 & 993.834$\pm$0.043 & 998.19\\
2 & 293.15 & 293.152 $\pm$0.010 & 989.386$\pm$0.042 & 997.30\\
3 & 297.95 & 297.949$\pm$0.005 & 986.872$\pm$ 0.033 & - -\\
4 & 303.05 & 303.050$\pm$0.006 & 984.104$\pm$0.029 & - -\\
5 & 303.15 & 303.152$\pm$0.003 & 984.051$\pm$ 0.031 & 995.61\\
6 & 308.25 & 308.242$\pm$0.001 & 981.093$\pm$ 0.053 & - -\\ 
7 & 313.15 & 313.153 $\pm$0.007 & 978.44  $\pm$ 0.059 & 994.20 \\
8 & 315.75 & 315.753$\pm$0.003 &  976.435$\pm$ 0.050 & - -\\
9 & 323.15 & 323.142 $\pm$0.003 & 971.565   $\pm$0.031  & 992.17\\ 
10 & 333.15 & 333.152 $\pm$0.006 & 964.396 $\pm$ 0.048  & 987.99\\ 
\hline
\end{tabular}}
\end{table}
\noindent Table (\ref{density_box}) shows that our simulated value of system density is in maximum deviation of around 1$\%$ with that of
water density. After equilibration run we perform the production run to calculate the equilibrium properties of the system such as diffusion
coefficient by fixing the number of particles, volume and temperature i.e.~ NVT ensemble. We use \emph{velocity-rescale} thermostat for this
case. We don't couple the system to a fixed pressure and use the structure obtained after equilibration run by which we fix the volume of
the system. The production run was carried out for 100 ns with the time step of 2 fs. 

\subsection{\small Energy Profile}
\begin{figure}[H]
\includegraphics[scale=0.38]{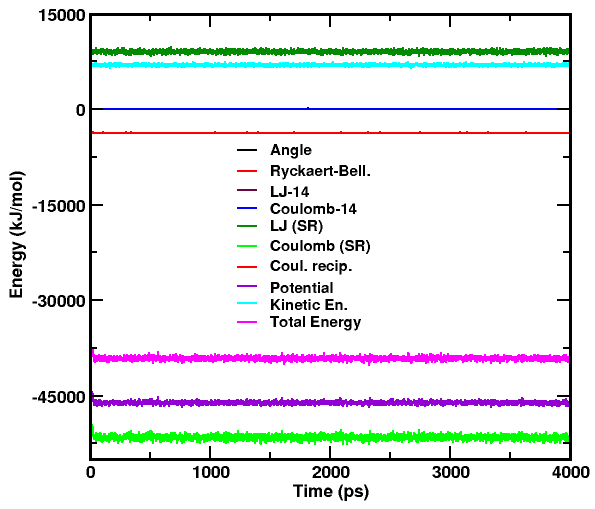}
\caption{Energy profile of the butane-water system at T=283.15 K.} \label{energy_283.15}
\end{figure}

\noindent  Figure (\ref{energy_283.15}) represents the energy profile of the butane-water system at $283.15$ K with the contributions of different energies. In our force field, total potential energy is the sum of bonded and non-bonded interaction energy. The bonded interaction includes bond stretching, bond angle vibration, proper dihedral, and improper dihedral. The non-bonded interaction includes LJ and coulomb interaction. Intra-molecular non-bonded interactions comprise Coulomb-14 and LJ-14 interactions. Energy profiles, which are variations of energy with time, are used to study the nature of these  Lennard-Jones and Coulomb energy. As we have used the cut-off values for Lennard-Jones and Coulomb potential the energy corresponding to them are the short range energies. The total energy is the sum of potential and kinetic
energies.
\noindent Looking at the energy profile we can observe that the Coulomb interaction (short range and reciprocal) and LJ interaction have significant contribution to the potential
energy of the system while the bonded interaction energies and the pair potential energies (14-interactions) contribute very less to the potential energy of the system. The Lennard-Jones interaction (LJ-SR and LJ-14) energy is positive  and the Coulomb (Coulomb-SR and Coulomb-Recip) energy is negative.
The bond angle, LJ-14, Coulomb-14 and Ryckart-Bellman dihedral energies are almost zero.  The total potential energy is$-46105.70\pm0.78$  kJ mol$^{-1}$ and the kinetic  energy is $6956.73\pm0.17$ kJ mol$^{-1}$. So the total energy, sum of potential and kinetic energy is $-39149.00\pm0.65$ kJ mol$^{-1}$. 
The negative value of the total energy shows that the system is bounded and is in stable equilibrium. 
\section{Results and Discussion} \label{results_discussions}
\noindent In this section, we  discuss the structural and dynamical properties  of the constituents of the  systems. 
\subsection{Radial Distribution Function}
\noindent Radial distribution functions (RDF) were obtained from the
simulations, in order to analyse the local structure around the solute and solvent molecule.  Radial distribution function (RDF) gives the 
idea of distribution of neighboring molecules with respect to the reference molecule considered in the calculations. In periodic systems, RDF
shows sharp peaks and troughs up to infinity where the separations and heights are the characteristics of the lattice structure. In liquids
however, RDF oscillates up to certain orders and then attains constant value as unity \cite{mcquarrie}.\\
\begin{figure*}[t]
\includegraphics[scale=0.40]{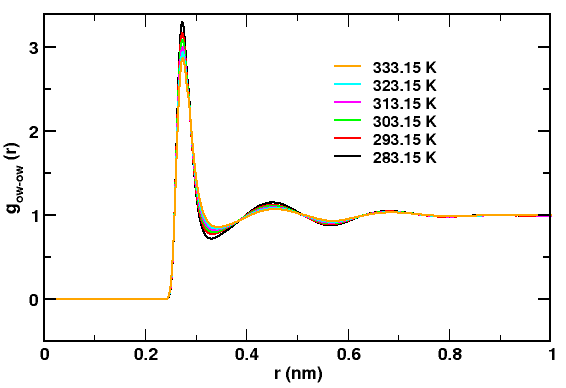} \hfill \includegraphics[scale=0.385]{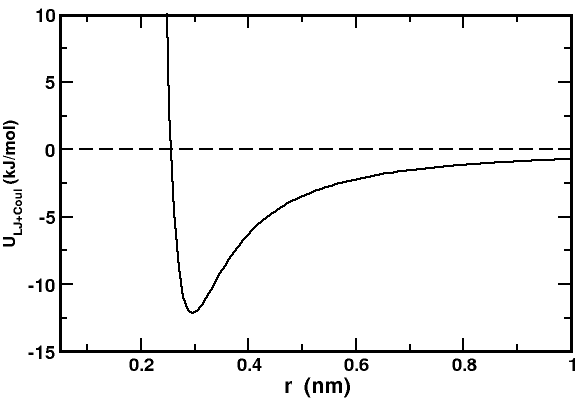}
\caption{(a) RDF of oxygen atoms of water molecules at different temperatures (left) (b)  Lennard-Jones plus Coulomb ($\mathrm{U_{LJ+Coul}}$) potential as a function of distance for two different isolated water molecules \cite{SP} (right)}
\label{rdfowow}
\end{figure*}
\noindent We have calculated RDF ${g(r)}$ of oxygen atoms of water molecules $g_{OW-OW}(r)$, oxygen of water and methyl ($\mathrm{CH_3}$) and methylene ($\mathrm{CH_2}$) carbons of  alkanes (methane, ethane, propane, butane) $g_{C-OW}(r)$.

\noindent Figure \ref{rdfowow}(a) represents the RDF of oxygen atoms of water molecules at different temperatures. For the structure  of the water molecule, the centre of mass is practically the same as the oxygen centre, which is also the van de Waals sphere centre. This makes the results for the oxygen atom representative of the whole water molecule. The figure explores three different peaks which implies that the molecules are correlated up to third solvation shell. The value of $\sigma$ for OW-OW is 0.3165 nm, and the van der Waals radius (2$^{\mathrm{1/6}}\sigma$) is 0.3553 nm \cite{gromacs user manual}. The figure \ref{rdfowow}(a) shows that excluded region remains fairly independent (0.276 $\pm$ 0.002 nm) of changing temperature. It also calculates that the excluded region is smaller than the van der Waals radius which indicates the contributions from other potentials in addition to the van der Waals potential \cite{SP} (see Fig.\ref{rdfowow}b). The first peak position remains at the same position within the error of $\pm$ 0.002 nm as a function of temperature. The magnitudes of all the peaks in RDFs decrease on rising temperature. Furthermore, the width of the peaks increases on increasing temperature. Both variations are the consequences of excess volume created in the system and  the  decrease in co-ordination number  with increase in temperature. These results show that the movement of the particles enhances and the solvent becomes less structured as temperature is increased. The  figures \ref{rdfowow} (a) and that  of  \ref{rdfowow} (b)\cite {SP} show that  Lennard-Jones plus Coulomb potential covers almost entire potential except many body effects. The second peak and third peak positions of the $g_{OW-OW}(r)$  are 0.450 $\pm$ 0.002 nm and 0.680 $\pm$ 0.002 nm respectively. These results are in good agreement with the available references \cite{pnas, exp}. From the simulations, we  found that the RDFs between oxygen atoms of water molecules in different alkane-water system are identical in all respects. It showed that the presence of the solute molecule has a negligible effect on the global structure of the solvent.\noindent
\begin{figure}
\includegraphics[scale=0.33]{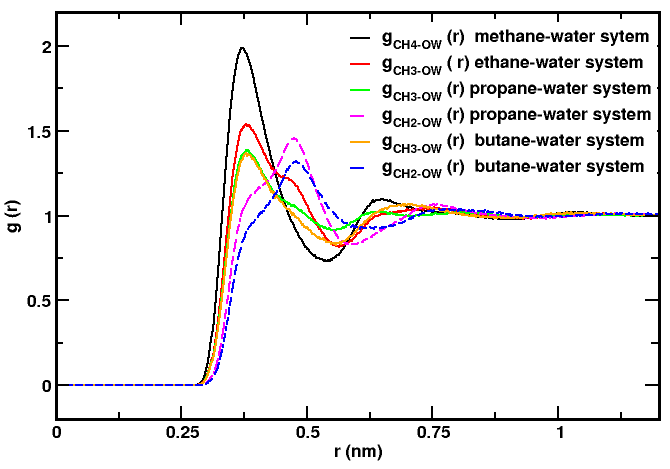}
\caption{Radial distribution functions between $\mathrm{CH_3}$ (solid
lines) and $\mathrm{CH_2}$ (dotted lines) and  oxygen atom of water molecule at 293.15 K.}
\label{rdfC-OW}
\end{figure}

The RDF between carbon of alkane and the oxygen of water describes solute-solvent interaction. Figure (\ref{rdfC-OW}) shows the RDF between the methyl ($\mathrm{CH_3}$) and methylene ($\mathrm{CH_2}$) carbons of the alkanes and  the oxygen atom of water, calculated from the simulations at 293.15 K. In figure (\ref{rdfC-OW}), it can be seen that height of the both $\mathrm{CH_3-OW}$  and $\mathrm{CH_2-OW}$ peak clearly decrease with increase in the length  of the carbon atoms  of the alkane. The methyl carbon  groups can always approach the water
molecule at closer distances (first peak position $ \sim 0.38$ nm), and the corresponding peaks are systematically
more intense than the $\mathrm{CH_2-OW}$ for distances under  $\sim 0.47$ nm.  Moreover, the magnitude as well as the excluded regions for $\mathrm{g_{CH3-OW}}$ and $\mathrm{g_{CH2-OW}}$ are different. This is because methyl  and methylene carbon do not possess the same partial charge. Furthermore, when oxygen of water (OW) approaches to methyl carbon, it (or the water molecule) also experiences the interactions due to three hydrogens attached to methyl carbon and when oxygen of water (OW)approaches to methylene carbon, it (or the water molecule) experiences the interactions due two hydrogens attached
to methylene carbon. This means when OW approaches to these carbons of alkane, it does not exactly experience the
same nature of interaction field around methyl and methylene carbon. The combination of these effects suggests that the methyl groups of alkane  molecules have a preferential tendency to be dissolved in the vicinity of water molecules and that this tendency decreases with chain length.  \\ 
\begin{figure}[H]
\includegraphics[scale=0.35]{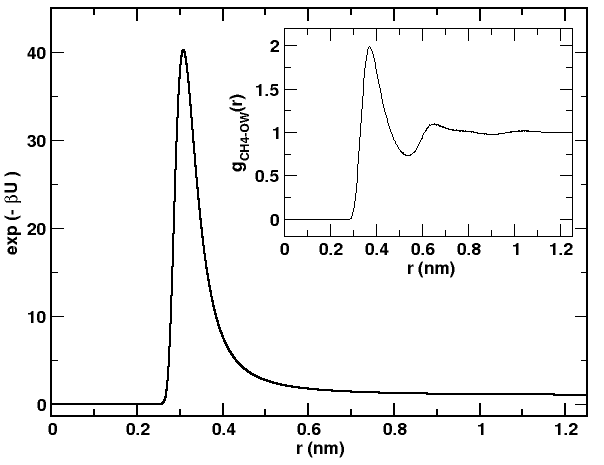}
\caption{Plot of exponential of negative of interaction potential between carbon atom of methane and water molecule and the corresponding radial distributaion function at 293.15 K. }
\label{interation}
\end{figure}
\noindent Fig.\ref{interation} represents the plot of exponential of negative of Lennard-Jones and Coulomb potential  between carbon atom of methane and water molecule as the function of  interatomic separation  and the corresponding radial distributaion function at 293.15 K. The maximum of $\mathrm{exp(-\beta  U)}$ (0.31 nm) and the first peak position (FPP) of the coreesponding radial distribution function (0.37 nm) are difffrent. This shows that   Lennard-Jones plus Coulomb potential doesn't cover almost entire potential.

\noindent Furthermore, to obtain the number of interaction sites ($N_c$) of each type in a coordination shell around the reference site,  we have integrated the
radial distribution functions (RDFs) as \cite{mcquarrie}:
\begin{equation}
N_c = \int_0^{r_{min}} 4\pi~ \rho~ g(r) r^2 dr
\end{equation}
Where $\mathrm{r_{min}}$ is the radius of the coordination shell  (location of the RDF minima) and $\rho$ is the number density. We have estimated the number of sites of a given groups or molecules around another groups or molecules, as a function of the distance from its centre. \\
\noindent In figure \ref{rdfowow}, for $g_{OW-OW}(r)$, the peak maxima ($\mathrm{r_{max}}$) of the first shell are obtained at $0.276$ nm  and the minima ($\mathrm{r_{min}}$) at $0.334$ nm for  all the alkane-water system. The first shell coordination number was
found to be $5.3\pm 0.1$ for water molecules. These values or the  coordination numbers are in good agreement  with the available reference values\cite{pnas, exp}. The first shell   co-ordination number of water molecules around  methyl carbon is $n_H\sim 23 $, in agreement with
the MAS NMR data\cite{NMR}. The first cell co-ordination number  of water for  methylene carbon is greater than that of methyl carbon. This result  also suggests that the methyl groups of alkane  molecules have a preferential tendency to be dissolved in the vicinity of water molecules. The details of the structural properties with the co-ordination numbers of water molecules around the methyl and methylene carbons of alkane-water system is provided in table \ref{coordination number}.
\begin{table}[H]
\caption{ Strucural paramerters from MD simulaion of alkane-water system at 293.15 K} 

\label{coordination number}
\resizebox {0.45 \textwidth } {!}{%
\begin{tabular}{c c c  c c} \hline
system & {groups}& $r_{max}$(nm) & ~~$r_{min}$(nm) &  CN
\\ \hline
CH4-H2O & CH4-water & 0.371 &  0.542  & 23.11\\   
C2H6-H2O & CH3-water & 0.378  & 0.565 & 25.49 \\  
C3H8-H2O & CH3-water & 0.381  & 0.555 & 23.18\\  
C3H8-H2O & CH2-water & 0.474  & 0.588 &28.52 \\
C4H10-H2O &  CH3-water & 0.378  & 0.558 & 22.22 \\ 
C4H10-H2O & CH2-water & 0.478  & 0.606 & 30.14 \\\hline
\end{tabular}} 
\end{table} 

 \subsection{Diffusion Coefficients}
\noindent The self-diffusion coefficient of alkane (methane, ethane, propane, butane) and water are calculated by using Einstein's relation (MSD method). 
\begin{figure}[H]
\includegraphics[scale=0.32]{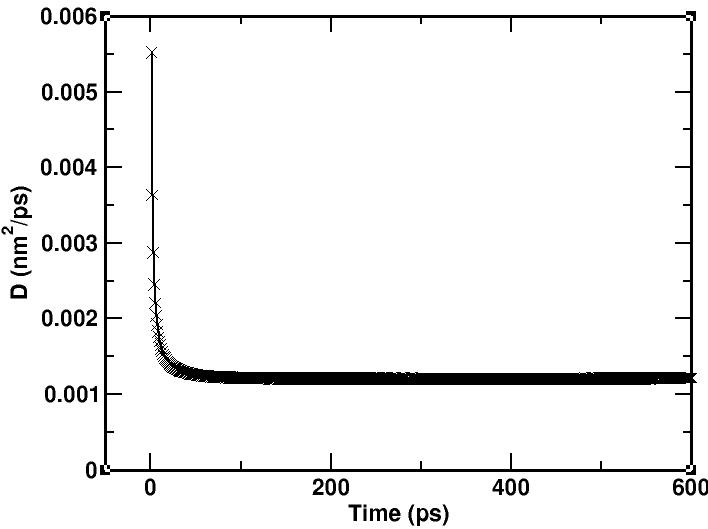}
\caption{Plot of Diffusion coefficient (D) = $\left\langle r^2(t)\right\rangle / 6\;t$ vs time of ethane at 283.15 K.}
\label{diffusionvstime}
\end{figure}
\noindent The figure (\ref{diffusionvstime}) shows the variation of diffusion coefficient (D) = $\left\langle r^2(t)\right\rangle / 6*t$  with time for ethane at temperature T =283.15 K. Although the production run was carried out for 100 ns, the plots of MSD curve
up to 2 ns for all temperatures are portrayed in the graphs. Reason for this is that the MSD curves show approximately a perfect linear nature up to 2 ns and calculated
self-diffusion coefficients are promising. Also, from figure (\ref{diffusionvstime}) we can say that it is reasonable to take time  2 ns while calculating diffusion coefficient as the graph is straight up to 2 ns. At first the diffusion coefficient is high due to ballistic motion and later as time passes it remains constant. This constant portion of the graph gives the diffusion coefficient \cite{SP}. 
\begin{figure}[H]
\includegraphics[scale=0.25]{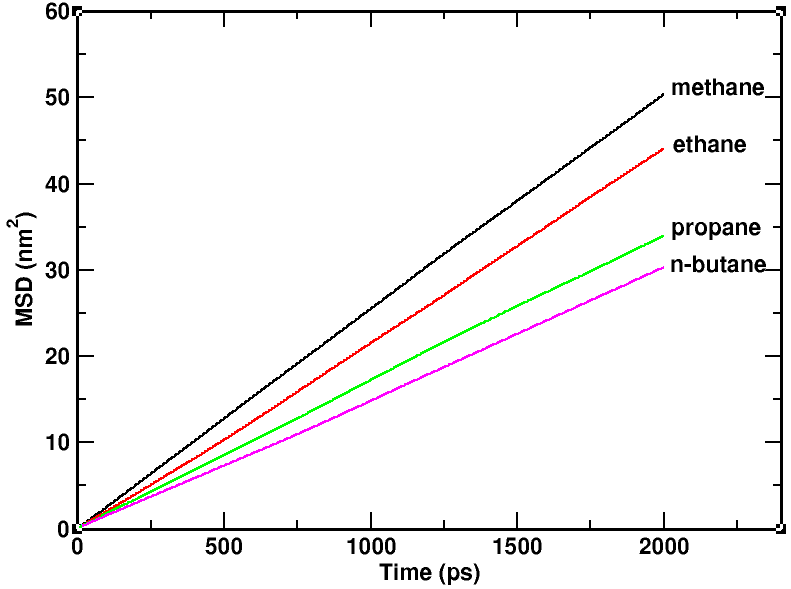}
\caption{Plot of MSD vs time of alkane at temperature 333.15 K.}
\label{msd}
\end{figure}
\begin{figure}[H]
\includegraphics[scale=0.40]{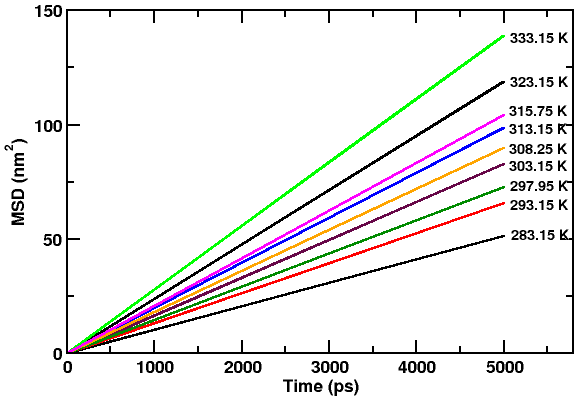}
\caption{Plot of MSD vs time of water at different temperatures.}
\label{msdh20}
\end{figure}
\noindent To obtain self-diffusion coefficient of alkanes and water via MSD we plot the graph between mean square displacement with time.
Figures (\ref{msd}, \ref{msdh20}) show the MSD plot of propane and water at different temperatures respectively. Since the statistics is better due to higher averaging at the starting than towards the ending region, we take the certain portion which has the best statistics. The value of self-diffusion coefficients of the desired species is calculated using equation  (\ref{MSD_equation}).  In our case, we have a simulation time of 100 ns and the best statistics for alkane (methane, ethane, propane and n-butane) molecule is found within 2 ns which can
also be justifiable from figure (\ref{diffusionvstime}) and is very small in comparison to simulation time this is due to lesser number of
alkane molecules. For water molecule best statistics is found within 5 ns due to larger number of water molecules. The binary diffusion coefficient of the alkane-water system is estimated using Darken’s relation (Eq.\ref{darken_equation}). Our system consists of 3 alkane molecules (methane, ethane, propane and n-butane each)  and 971 water molecules, a separate sytem, so the mole fraction for alkane is 0.003 and that of water is water is 0.997. The binary diffusion coefficient is  very close to that of self-diffusion coefficient of solute in the mixture due to low solute concentrations studied in this work. The values of self diffusion coefficients of water ($\mathrm{H_2O}$) at different temperatures is presented in the table \ref{water_diffusion}.
\begin{table}[h]
\caption{ Self-Diffusion Coefficients of water at different temperatures and the reference experimental values } 
\label{water_diffusion}
\resizebox {0.45 \textwidth } {!}{%
\begin{tabular}{   c| c  c } \hline
Temperature  & \multicolumn{2}{c} {Diffusion Coefficients ( $\times$10$^{-\mathrm{9}}$ m$^\mathrm{2}$s$^{-\mathrm{1}}$)} 
\\ (K) & Simulated Value & Experimental Value\cite{water_experimental1} \\ \hline
283.15 & 1.71 & 1.54  \\   
293.15 & 2.18 & 2.02    \\  
297.95 & 2.43 & -- \\  
303.05 & 2.70 & --   \\
303.15 &  2.73 & 2.59    \\
308.25 &  3.01 & --   \\
313.15 &  3.30 & 3.24    \\ 
315.75 &  3.46 & --   \\
323.15 &  3.95 & 3.96   \\
333.15 & 4.64 & 4.77    \\\hline
\end{tabular}} 
\end{table} 

 The values of the self-diffusion coefficient of alkane (methane, ethane, propane, n-butane) and  water obtained from MSD plot and the binary diffusion coefficients along with the references at different temperatures   are presented in table (\ref{diffusion_table}).
 \begin{table}[H]
\caption{The simulated value of binary diffusion coefficient of alkanes (methane, ethane, propane, n-butane)  and also the references for them 
 as a function of temperature are listed.}
\label{diffusion_table}
\resizebox {0.48 \textwidth }{!}{%
\begin{tabular}{c c c c c }
\hline 
\multicolumn{5}{c}{Diffusion Coefficient  ( $\times$10$^{-\mathrm{9}}$ m$^\mathrm{2}$s$^{-\mathrm{1}}$)} \\  
\hline  \\

System & Temp.(K) &  Simulation & Experimental (1)\cite{Houghton} & Experimental (2)\cite {Witherspoon} \\ \\ \hline
 &  283.15 & 1.79 & 1.9 & ---    \\
 & 293.15 & 2.08 & 2.4 & ---    \\
 & 297.95 & 2.35 & --- & 1.88    \\
 & 303.05 & 2.56 & --- & ---   \\
CH4-H20 & 303.15 & 2.60 & 3.0 & --- \\
 & 308.25 & 2.80 & --- & 2.12    \\
 & 313.15 & 3.34 & 4.2 & ---    \\
 & 315.75 & 3.50 & --- & 2.41   \\
 & 323.15 & 3.96 & 4.7 & ---   \\
 & 333.15 & 4.36 & 6.7 & ---   \\ \hline & 283.15 & 1.29 & 1.6 & ---    \\
 & 293.15 & 1.54 & 2.3 & ---    \\
 & 297.95 & 1.66 & --- & 1.52    \\
 & 303.05 & 1.89 & --- & 1.59    \\
 C2H6-H20& 303.15 & 1.90 & 2.8 & --- \\
 & 308.25 & 2.18 & --- & 1.72   \\
 & 313.15 & 2.32 & 3.3 & ---    \\
 & 315.75 & 2.45 & --- & 1.95    \\
 & 323.15 & 2.84 & 4.1 & ---    \\
 & 333.15 & 3.61 & 4.9 & ---    \\
 \hline & 283.15 & 0.93 & 1.3 & --- \\
 & 293.15 & 1.23 & 1.8 & ---    \\
 & 297.95 & 1.41 & --- & 1.21    \\
 & 303.05 & 1.60 & --- & 1.27    \\
C3H8-H20 & 303.15 & 1.65 & 2.4 & ---    \\
 & 308.25 & 1.81 & --- & 1.39   \\
 & 313.15 & 1.97 & 2.7 & ---    \\
 & 315.75 & 2.10 & --- & 1.59    \\
 & 323.15 & 2.53 & 3.5 & ---    \\
 & 333.15 & 2.85 & 4.4 & ---    \\
 \hline
 & 283.15 & 0.84 & 0.83 & ---    \\
 & 293.15 & 1.09 & 1.4 & ---   \\
 & 297.95 & 1.26 & --- & 0.96    \\
 & 303.05 & 1.39 & --- & 1.03    \\
C4H10-H20 & 303.15 & 1.41 & 1.9 & ---    \\
 & 308.25 & 1.70 & --- & 1.12    \\
 & 313.15 & 1.90 & 2.5 & ---    \\
 & 315.75 & 2.00 & --- & 1.28  \\
 & 323.15 & 2.25 & 3.3 & ---    \\
 & 333.15 & 2.58 & 4.3 & ---    \\
\hline 
\end{tabular}}
\end{table} 
\begin{figure}[H]
\includegraphics[scale=0.28]{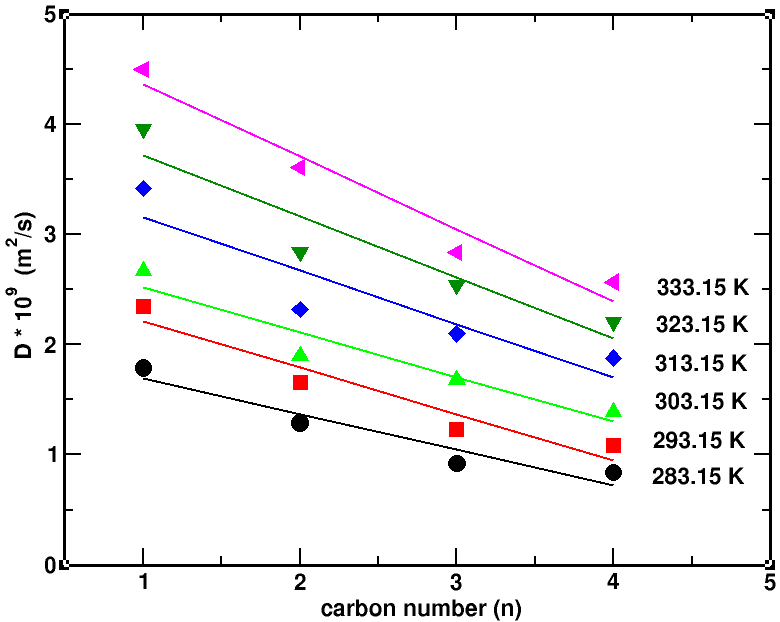}
\caption{The variation of simulated values of binary diffusion coefficinets  of alkanes in water  with the numbers of carbon atoms in the alkane chain  at different temperatures } 
\label{carbon}
\end{figure}
\noindent Figure (\ref{carbon}) shows the variation of binary diffusion coefficients of the molecule with increasing the numbers of carbon atoms of the alkane chain. The diffusion coefficents of the molecules decreases with increasing the number of carbons present in the alkane molecules. Thus, the diffusion coefficients of methane is highest and that of n-butane is lowest.  
\begin{figure}[H]
\includegraphics[scale=0.40]{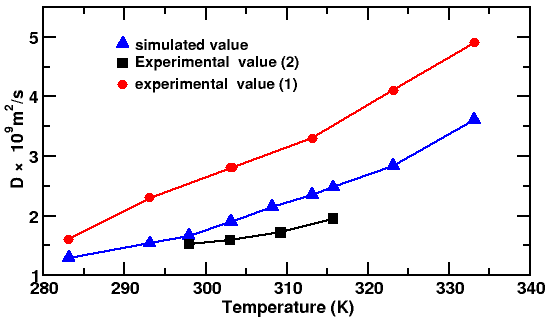}
\caption{Simulated and experimental values of binary diffusion coefficients of ethane-water system at different temperatures}
\label{compare}
\end{figure}

\noindent Table (\ref{diffusion_table}) shows the self and binary diffusion coefficients of alkane and H$_\mathrm{2}$O molecules from the present work along with the references at different temperatures. The comparison of the values from the table and also from other references explores that self-diffusion coefficients of water from the present work, in general, come in very good agreement with the previous studies \cite{water_experimental1, sunil's paper, Poudyal2014, SP}. 
The experimental and simulated values of self-diffusion coefficients of water in all system are in good agreement with maximum  deviation of $11\%$  at 283.15 K  \cite{water_experimental1}. The simulated values of alkane, on the other hands, show different attitude towards the references. They lie very well in between the experiment performed by (1) D. L. Wise and G. Houghton\cite{Houghton}
and (2) P. A. Witherspoon and D. N.
Saraf \cite {Witherspoon}. Figure \ref{compare} is the comparision of simulated values with the experimental values \cite{Houghton, Witherspoon} of binary diffusion coefficients of ethane-water system at different temperatures. They lie very well in between the experimental values\cite{Houghton, Witherspoon}  within the error of $33\%$. The deviations of the simulated values with the experimental values follows the same trends in all alkane-water system.    There are very large differences in the values  of the binary diffusion coefficients reported by them \cite{Houghton, Witherspoon}. The diffusion coefficient for both the solute (alkane) and solvent (water)
molecules increases with the enhanced temperature, which is due to the increase in the velocity of the molecules, as per relation of the
thermal energy with temperature. Furthermore, as the density of the system decreases with increasing temperature, the space available for the alkane molecules to execute random-walk motion increases\cite{sunil's paper}. Finally, based on these facts, the mean squared displacement
increases and this change is incorporated by Einstein’s relation to yield an increased self-diffusion coefficient.
\subsection{Temperature dependence}
\begin{figure}[H]
\includegraphics[scale=0.35]{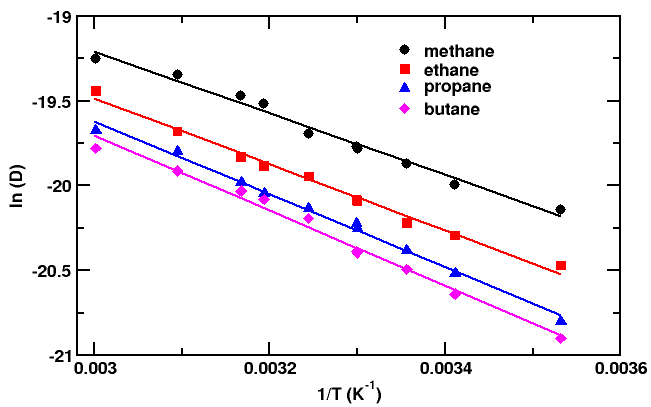}
\caption{Arrhenius diagram of the simulated values of binary diffusion coefficients of alkane in water.}
\label{arrhenius_alkane}
\end{figure}

\begin{figure}[H]
\includegraphics[scale=0.41]{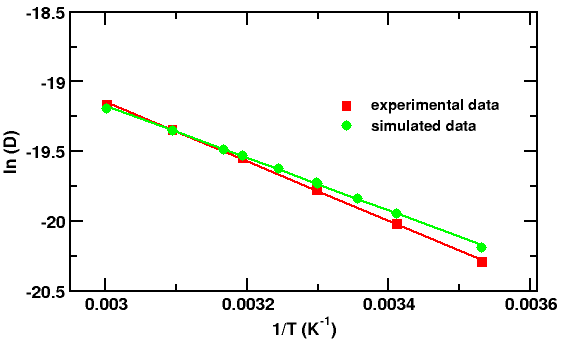}
\caption{Arrhenius diagram of the simulated and experimental value of water.}
\label{arrhenius_water}
\end{figure}

 \noindent Diffusion coefficients of a system generally depends strongly on temperature, being low at low temperatures and are found to be increased with increase in temperature. Temperature variations in diffusivity is explained by the Arrhenius formula  \cite{diffusion in solids}.
\begin{equation} \label{Arrhenius_equation}
\mathrm{D}= \mathrm{D}_0~\mathrm{exp}\left(-E_a/R \; T\right)
\end{equation}
which can be expressed as 
\begin{equation}
ln D = ln D_0 - E_a/R\;T
\label{arrentype}
\end{equation}
where, D$_0$ denotes the pre-exponential factor, also called frequency factor, E$_a$ is the activation energy for diffusion, $T$ is the absolute temperature, R =N$_A$ k$_B$ is molar gas constant whose value is is $8.31~ \mathrm{J ~{mol}^{-1} K^{-1}}$. Both E$_a$ and D$_0$ are called the activation parameters of diffusion. The simulated binary diffusivities of 
table (\ref{diffusion_table}) have been fitted to Arrhenius-type expression equation (\ref{arrentype}) by least
squares method, the pre-exponential constant $\mathrm{D_0}$ and the activation energy $\mathrm{E_a}$ are reported in table (\ref{activation_energy}).  
\begin{table}[H]
\caption{Table for Activation energies and pre-exponential factors for diffusion of various studied system 
.} 
\label{activation_energy}
\resizebox {0.45 \textwidth } {!}{%
\begin{tabular}{c  c  c } \hline \\
\multicolumn{3}{c} 
{system} ~~~~$E_a$ (kJ $mol^{-1}$)  ~~~ ~~$D_0 \times 10^7 m^2/s $  \\
\\ \hline
CH4-H2O & 15.13 &  7.00\\  
C2H6-H2O & 16.18  & 11.86 \\  
C3H8-H2O & 17.81  & 18.67 \\  
C4H10-H2O & 18.38  & 22.63 \\  
H2O & 15.59  & 13.12 \\\hline
\end{tabular}} 
\end{table} 
\noindent Figure(\ref{arrhenius_alkane})  shows the temperature dependence of diffusion coefficient of alkane in water. As the simulation  data fit to the equation (\ref{Arrhenius_equation}), the temperature dependence of diffusion coefficient follows Arrhenius
 behavior. From figure (\ref{arrhenius_alkane}), it is seen that the diffusion coefficients increase with increase in temperatures. This could be due to the fact that at higher temperature the difference in the density of the system (i.e. alkane and water) and water increases with
 increase in temperature (see table (\ref{density_box})).
\noindent Figure (\ref{arrhenius_water}) shows the temperature dependence of diffusion coefficient of water in the system containing water
and alkane. Figure (\ref{arrhenius_water}) explicitly shows the temperature dependence of diffusion coefficient of water also follows Arrhenius behavior.

\section{Conclusions and Concluding Remarks} \label{conclusions}
\noindent In this work, we have computed self diffusion coefficients along with binary diffusion coefficients of the system containing 971 water (H$_\mathrm{2}$O) molecules and 3 alkane (methane, ethane, propane, n-butane) molecules over a wide range of  temperatures from 283.15 K - 333.15 K,  using molecular dynamics simulation technique. The Extended Simple Point Charge (SPC/E)  model of water and Optimized Potential for Liquid Simulations- All Atom (OPLS-AA) of alkane were used. Here alkane molecule acts as a solute and water
(H$_\mathrm{2}$O) as a solvent.  The energy profile (figure \ref{energy_283.15}) of the system were studied to know the equilibrium nature of
the system. \\ 
\noindent  Structural properties has been studied using Radial Distribution Function (RDF) and co-ordination numbers of the interaction cites has been calcualted integrating RDF to the first co-ordination shell. The obtaind RDFs  show that the system becomes less structured at high temperatures. The equilibrium structural properties of both the components (alkane and water) were studied calculating
corresponding radial distribution function (RDF) namely $g_{OW-OW}(r)$ RDF of oxygen atoms of water molecules, $g_{CH_3-OW}(r)$ RDF of carbon atom of methyl group of alkane and oxygen atom of H$_\mathrm{2}$O , $g_{CH_2-OW}(r)$ RDF of carbon atom of methylene group of alkane and  oxygen atom of H$_\mathrm{2}$O. \\
\noindent The main aim of our work was to study diffusion phenomenon of the mixture of water and alkane and study its temperature dependence. The self-diffusion coefficients of water and alkane (methane, ethane, propane and n-butane) was estimated using Einstein's method  separately. The diffusion coefficients of water are deviated within 11$\%$ of the available experimental data \cite{water_experimental1}. The binary  diffusion coefficient of the system was calculated using Darken's relation. The values of binary diffusion coefficients of alkane in water do not agree well with the experimental values of D. L. Wise and G. Houghton\cite{Houghton}
and  P. A. Witherspoon and D. N.
Saraf \cite {Witherspoon}. It lies in between these two experimental values  and the deviation is increasing with increase in temperature. The Arrhenius diagram (plot of natural logarithm of diffusion coefficient vs inverse of temperature) was plotted for self-diffusion coefficients of water and binary diffusion of the alkane-water system  separately and it showed temperature dependence of diffusion coefficient of both are of Arrhenius type.\\ 
\begin{center}
\textbf{ACKNOWLEDGEMENTS}
\end{center}\vspace*{0.4cm}
\noindent One of the autors S. Pokharel  acknowledges	 the partial support from the Abdus Salam	International Centre	for	 Theoretical Physics,	
  Trieste, Italy through the office of external activities (OEA). Further, A.~ Subedi, N.~ Aryal and B.~R.~ Niraula acknowledge the partial financial support from University Grants Commission (UGC), Nepal. 

\end{document}